\documentclass[a4paper]{aa}
\usepackage{graphicx,float}
\usepackage{latexsym,amsmath,amssymb}
\usepackage{natbib}
\usepackage{changebar}
\usepackage[scriptsize,tight]{subfigure}
\bibpunct{(}{)}{;}{a}{}{,}
\def\integral{{\it INTEGRAL}}

\begin{document}

\title{The hard X-ray view of bright infrared galaxies}

\author{R. Walter\inst{1, 2} \and N. Cabral\inst{1,2,3}}
\institute{
INTEGRAL Science Data Centre, Chemin d'Ecogia 16, CH-1290 Versoix, Switzerland, \email{Roland.Walter@unige.ch}
\and
Observatoire de Gen\`eve, Universit\'e de Gen\`eve, Chemin des Maillettes 51, CH-1290 Sauverny, Switzerland
\and
Universit\'e de Bordeaux 1, UFR de Physique, 351 Cours de la Lib\'eration, F-33405 Talence Cedex, France
}
\offprints{Roland Walter}

\date{Received 29 August 2008 / Accepted 30 January 2009}

\abstract{}{The synthesis of the cosmic X-ray background (CXB) requires a large population of Compton-thick active galactic nuclei that have not been detected so far. We probe whether bright infrared galaxies could harbor a population of Compton-thick nuclei and if they could contribute significantly.}
{We analyzed 112 Msec of \integral\ observations obtained on 613 galaxies from the IRAS Revised Bright Galaxy Sample. 
We derived the average hard X-ray (18-80 keV) emission of Seyfert and various non Seyfert galaxy subsamples to estimate their relative contribution to the locally emitted CXB.}
{The Seyfert 1 \& 2 are detected at hard X-rays. None of the other galaxy subsamples were detected. ULIRGs are at least 5 times under-luminous at hard X-rays when compared to Seyferts. The upper limit obtained for the average non Seyfert galaxies is as low as $7\times 10^{-13}$ erg s$^{-1}$ cm$^{-2}$. On average, these galaxies do not contain active nuclei brighter than $10^{41}~{\rm erg/s}$ at hard X-rays.
The total hard X-ray flux detected from the sample is $4.9\times 10^{-9}$ erg s$^{-1}$ cm$^{-2}$ (about 1\% of the CXB), and 64\% of this originates in absorbed active nuclei. Local non-Seyfert galaxies contribute for less than 7\% and do not harbor the Compton-thick nuclei assumed to synthesize the locally emitted CXB.}{}


\maketitle

\section{Introduction}
\label{sec:intro}

While accreting super-massive black-holes are believed to power active galactic nuclei \citep[AGN; ][]{Lynden-Bell1969}, the apparent spectral energy distribution of these objects
depends on many physical parameters, such as the black-hole mass, the accretion rate, the nature of the accretion flow, the inclination on the line of sight, 
and the intrinsic obscuration. 

The hard X-ray background, with a prominent maximum around 30 keV, requires that most AGNs are obscured  \citep{Setti1989}. Quantitatively, population synthesis models indicate that 60 to 80\% of all AGNs should be obscured and that about 50\% of these should be Compton-thick \citep{Gilli2007}, i.e. featuring absorbing column densities $N_{\rm H}\ge \sigma_{\rm T}^{-1} \approx 1.6\times10^{24}~{\rm cm}^{-2}$.
Nuclear starbursts in the central few parsecs of a galaxy could provide the necessary obscuring material \citep{Ballantyne2008} to explain a large population
of Compton-thick objects. These nuclear starbursts also link stellar formation and the black-hole activity \citep{Watabe2008, Kawakatu2008}.

Various techniques have been used to study the relation between starbursts and AGN activity, based on the study of the infrared spectral energy 
distribution \citep{Genzel1998} or on correlations between infrared and X-ray or radio detections \citep{Barger2005, Polletta2006}. 
Because hard X-ray emission is an ubiquitous signature of AGN activity \citep{Mushotzky1993} and as it is insensitive to photoelectric absorption, the correlation 
between infrared and hard X-ray fluxes is an interesting probe for the existence of highly absorbed AGNs. 

Even if a fraction of the AGN population features various signatures of absorption -- like an absence of broad emission lines, scattered emission \citep{Schmidt2007}, 
reddening \citep{WalterFink1993, Wilkes2005}, or soft X-ray photoelectric absorption \citep{Tozzi2006} --
recent low sensitivity hard X-ray surveys \citep{Paltani2008,Tueller2008} have failed to detect the population of Compton-thick objects detected at low X-ray fluxes \citep{Risaliti1999, Guainazzi2005}.

To improve the sensitivity of the current hard X-ray surveys by one order of magnitude, we stacked the hard X-ray signal measured by \integral\
for various types of infrared bright galaxies to check their contribution to the hard X-ray background. The sample of sources is presented in section \ref{sec:sample}, the \integral\ data and analysis is described in section \ref{sec:integral} and their correlation with infrared data 
is discussed in section \ref{sec:discussion}.

\section{Source sample}
\label{sec:sample}

In 1983, the Infra-Red Astronomical Satellite (IRAS) surveyed approximately 96\% of the sky in four broad wavelength bands centered on 12, 25, 60 and 100 $\mu$m. The IRAS Point Source Catalog (PSC) and the IRAS Faint Source Catalog (FSC) were produced listing 245889 and 173044 infrared sources respectively (the PSC extends to lower galactic latitude than the FSC). IRAS discovered the luminous and ultra-luminous  infrared galaxies (LIRGs and ULIRGs) in the local Universe with infrared luminosities larger than 10$^{11}~{\rm L}_{\odot}$ and 10$^{12}~{\rm L}_{\odot}$ respectively.

\cite {Sanders2003} used the IRAS FSC and PSC to constitute the IRAS Revised Bright Galaxy Sample (RBGS). The RBGS is a complete flux-limited sample of all extragalactic objects brighter than 5.24 Jy at 60 $\mu$m, covering the entire sky surveyed by IRAS at galactic latitudes $\vert{\rm b}\vert>5\degr$. 
The median redshift of the RBGS galaxies is z=0.0082 and their bolometric luminosity function breaks at $\sim10^{10.5}L_{\odot}$ from a slope of $-0.6$ to $-2.2$.

Among the 629 objects of the RBGS, we excluded  (Table~\ref{tab:isolatedsources}) the SMC and LMC because they are extended for the imager on board \integral, 13 objects because of their proximity with other sources detected at hard X-rays and NGC 6240, one of the rare object hosting two massive black-holes. NGC 6240 is an ULIRG and includes a starburst, a Compton-thick and a normal Seyfert nuclei \citep{Risalitti2006}. The origin of the hard X-ray emission \citep{Vignati1999} detected by \integral\ at a level of 1 mCrab is therefore unclear.

The resulting sample of 613 objects was classified in six sub-samples according to their object type defined by SIMBAD\footnote{http://simbad.u-strasbg.fr/}:  93 Seyfert galaxies (including 14 Seyfert 1, 64 Seyfert 2 and 15 Seyfert of unclassified or complex types) and 520 ``non Seyfert" galaxies, missing clear signature for an active nucleus, including 12 ULIRGs $({\rm L}_{60{\rm \mu m}}/{\rm L}_{\odot}>10^{12})$, 144 LIRGs $(10^{12}>{\rm L}_{60{\rm \mu m}}/{\rm L}_{\odot}>10^{11})$ and 41 Liners. 
Among the 14 Seyfert 1 galaxies of our sample, several objects have in fact intermediate Seyfert types. 
NGC 1386, NGC 2992, NGC 5033 are Seyfert 1.9 but their X-ray spectra do not show significant absorption \citep{Dadina2008,Terashima1999}. 
NGC 3227 and NGC 4151 are Seyfert 1.5 with complex and variable absorption \citep{Gondoin2003,Zdziarski2002}. The division of the Seyfert 
sample according to their Seyfert type is only an approximation of their absorption properties in the X-rays \citep[e.g.][]{Bassani2006}.

\begin{table}[t]
\caption{Sources from the RBGS excluded from the analysis and reason for their exclusion.}
\begin{tabular}{p{5cm}p{3cm}}
\hline \hline \noalign{\smallskip}
Sources& Reason\\ 
\noalign{\smallskip\hrule\smallskip}
NGC 2339, IRAS 05405+0035, NGC 2342, IRAS 05442+1732, IRAS 05223+1908 & Close to Crab\\
\noalign{\medskip}
NGC 2993, NGC 3583, NGC 4402, NGC 4303, NGC 6215, NGC 7590, NGC 7599, IRAS 18293-3413 & Close to other bright sources\\
\noalign{\medskip}
LMC, SMC & Extended\\
\noalign{\medskip}
NGC 6240  & Detected, complex\\
\noalign{\smallskip\hrule\smallskip}
\end{tabular} 
\label{tab:isolatedsources}
\end{table}

\section{INTEGRAL data and analysis}
\label{sec:integral}

All public data obtained by the \integral\ soft $\gamma$-ray imager {\it IBIS/ISGRI} \citep{ubertini03AA, lebrun03AA} until January 2007 are considered in this study. The \integral\ imager has a large field of view of 29$\degr$ square with a spatial resolution of 12 arcmin. Several thousands pointings, with typical exposures of $(1-3)\times 10^3$ s, including at least one source of the sample in its field of view and with an effective exposure longer than 120 s were selected, spanning times between Dec.~30, 2002 (revolution 26) and Jan.~17, 2007 (revolution 520).

The {\it ISGRI} data were reduced using the \integral\ Offline Scientific Analysis software\footnote{http://isdc.unige.ch/} version 7.0 publicly released by the \integral\ Science Data Centre \citep{courvoisier03AA}. Individual sky images for each pointing have been produced in a broad energy band of 18--80~keV. All known ISGRI sources, with significance above $5\sigma$, have been used for image cleaning. The position of these sources was fixed to catalogued values when known accurately. 

To measure the average hard X-ray properties of the selected galaxy subsamples we have built mosaic images of all subsample sources modifying the coordinate system of each individual image such that the coordinates of each subsample sources were set to an arbitrary fixed position ($\alpha$=0, $\delta$=0). The resulting image provides a stack of all {\it ISGRI} data available on all sources of interest.

To minimize the systematics in the mosaic images, we have excluded 1455 individual sky images (many obtained before revolution 38, when the IBIS bottom anti-coincidence was reconfigured) which had background fluctuation rms larger than $1.1\sigma$ in the significance image and 779 images with minimum significance lower than $-5.5\sigma$. 36030 images were finally included in the processing (some of them many times, when including several sources).

\begin{figure}[t]
\centerline{\includegraphics[angle=0,width=7cm]{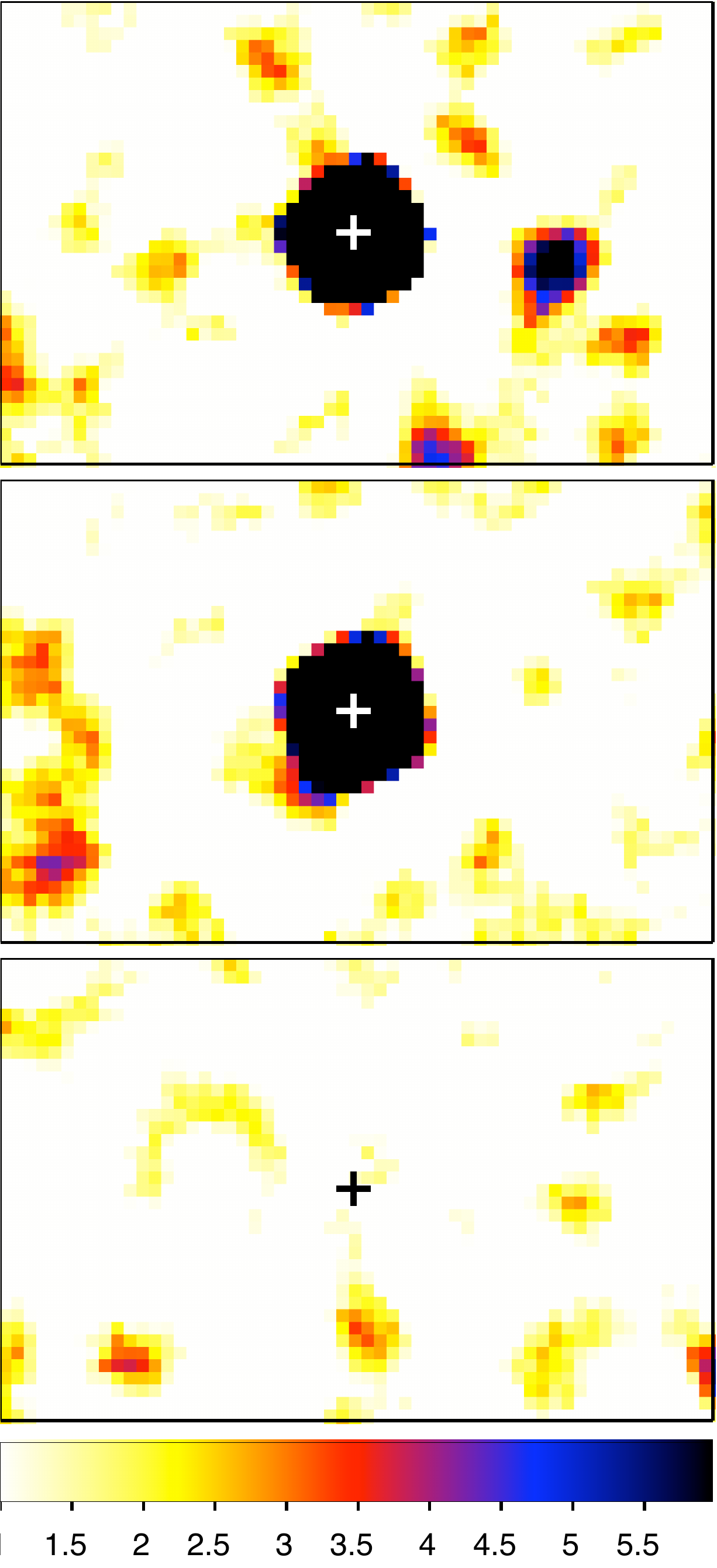}}
\caption{18-80~keV stacked mosaic significance images for three of the various galaxy subsamples. From top to bottom: Seyfert 1, Seyfert 2 and ``non Seyfert" subsamples. The color map spans significances between 1 and 6. The color map of the bottom image has been slightly modified to account for the systematics, re-normalizing the centroid of the background distribution to a value of 0.}
\label{fig:mosa}
\end{figure}

The 500x500-pixels mosaic images were built in equatorial coordinates centerred on ($\alpha$=0, $\delta$=0) with a tangential projection using a factor two over-sampling when compared to the individual input sky images; this results in a pixel size of 2.4 arcmin in the center of the mosaic. The photometric integrity and accurate astrometry are obtained by calculating the intersection between input and output pixels and weighting count rates according to the overlapping area. A composite of the central part of three mosaic images is featured in Fig. \ref{fig:mosa}.

The effective exposure obtained at the center of the mosaics are between 1.7 and 96.0 Ms for the various galaxy subsamples. To investigate the quality of the mosaics, we studied the distribution of the pixelsÕ significance, which is expected to be Gaussian with average 0 and dispersion 1. We fitted a Gaussian distribution to the part of the histogram with $\sigma\leqslant 3$ to avoid the strong positive tail due to real sources. While the width of the pixel distributions ($\approx 1.2$) are not unusual the centroid of the distribution is significantly displaced towards positive values for the sample of ``non Seyfert" galaxies where the effective exposure reaches almost 100 Ms. Our analysis shows that the individual {\it ISGRI} images appear slightly positive on average (0.004 ct/s) which could not be detected with typical exposures up to several Ms. Even if the exact cause of this effect is not known (background subtraction ?), this effect can be corrected for by subtracting a constant value to the mosaic image. 

A positive average signal is clearly detected in the mosaics built for the three Seyfert subsamples. No average excess is detected for the other galaxy subsamples. As expected, the position ($\alpha$=0, $\delta$=0) is well within the 90\% error box corresponding to the excesses found for the three Seyfert subsamples and the $1\sigma$ width of these excesses are between 5.7 and 6.0 arcmin, in good agreement with the ISGRI point spread function.

Table \ref{tab:flux} lists the average signal (or $5 \sigma$ upper limits) extracted at hard X-rays from each {\it ISGRI} mosaic with {\tt mosaic\_spec} (corrected for the higher sky background systematics obtained for the large ``non Seyfert" galaxy subsample) together with the average flux measured by IRAS at 60 $\mu$m \citep{Sanders2003} for each subsample. The hard X-ray fluxes were derived from the count rates using a spectral index of $\Gamma=1.8$. The hard X-ray detections obtained for the three Seyfert subsamples are more than one order of magnitude brighter than the upper-limits obtained for the other galaxy subsamples with comparable effective exposures. The hard X-ray upper limit obtained for the ``non Seyfert'' subsample is a hundred times less than the flux detected in average from the Seyfert galaxies. 

At hard X-rays, NGC 4151 dominates the Seyfert 1 subsample. Excluding it reduces the average count rate from 2.25 to 0.51.
The average count rate obtained for the small Seyfert 1 subsample is significantly affected by NGC 4151, contrasting with the average 60 $\mu$m flux that is almost not affected.

\begin{table}[t]
\caption{Average signals (or $5\sigma$ upper limits) detected for the various subsamples at hard X-rays (18--80 keV) and at 60 $\mu$m. Effective hard X-ray exposure, count rates and integrated flux are provided together with the average 60 $\mu$m flux. The hard X-ray detection significances for the first four samples are 119, 23, 122 and 147 respectively. }
\label{tab:flux}
\begin{tabular}[c]{lrrccc}
\hline \hline \noalign{\smallskip}
Sample&Srcs&Exp.&Ct rate&F$_{\rm 18-80keV}$&F$_{\rm 60\mu m}$\\
               &&{\tiny Ms}&{\tiny ct/s}&{\tiny $10^{-10}$ cgs}&{\tiny Jy}\\
\noalign{\smallskip\hrule\smallskip}
Seyfert 1      &14&1.7     &2.25$\pm$0.02&$1.28$&18.3\\
S1 - 4151\footnotemark[1]&13&1.4     &0.51$\pm$0.02&$0.29$&19.3\\   
Seyfert 2      &64&12.1   &0.87$\pm$0.01&$0.49$&31.1\\
Seyfert         &93&16.1   &1.04$\pm$0.01&$0.59$&28.9\\      
\noalign{\smallskip\hrule\smallskip}
ULIRGs       &12& 4.1     &$<0.06$&$<0.034$&10.2\\
Liners           &41& 7.5     &$<0.045$&$<0.026$&29.6\\
LIRGs          &144&26.7    &$<0.02$&$<0.011$&12.9\\
non Seyfert&520&96.0    &$<0.013$&$<0.0074$&19.1\\
\noalign{\smallskip\hrule\smallskip}
\end{tabular}
\footnotemark[1]includes all Seyfert 1 galaxies excepting NGC 4151.
\end{table}

\section{Discussion}
\label{sec:discussion}

\subsection{Average spectral energy distribution}

\begin{figure}[h]
\centerline{\includegraphics[angle=0,width=8.5cm]{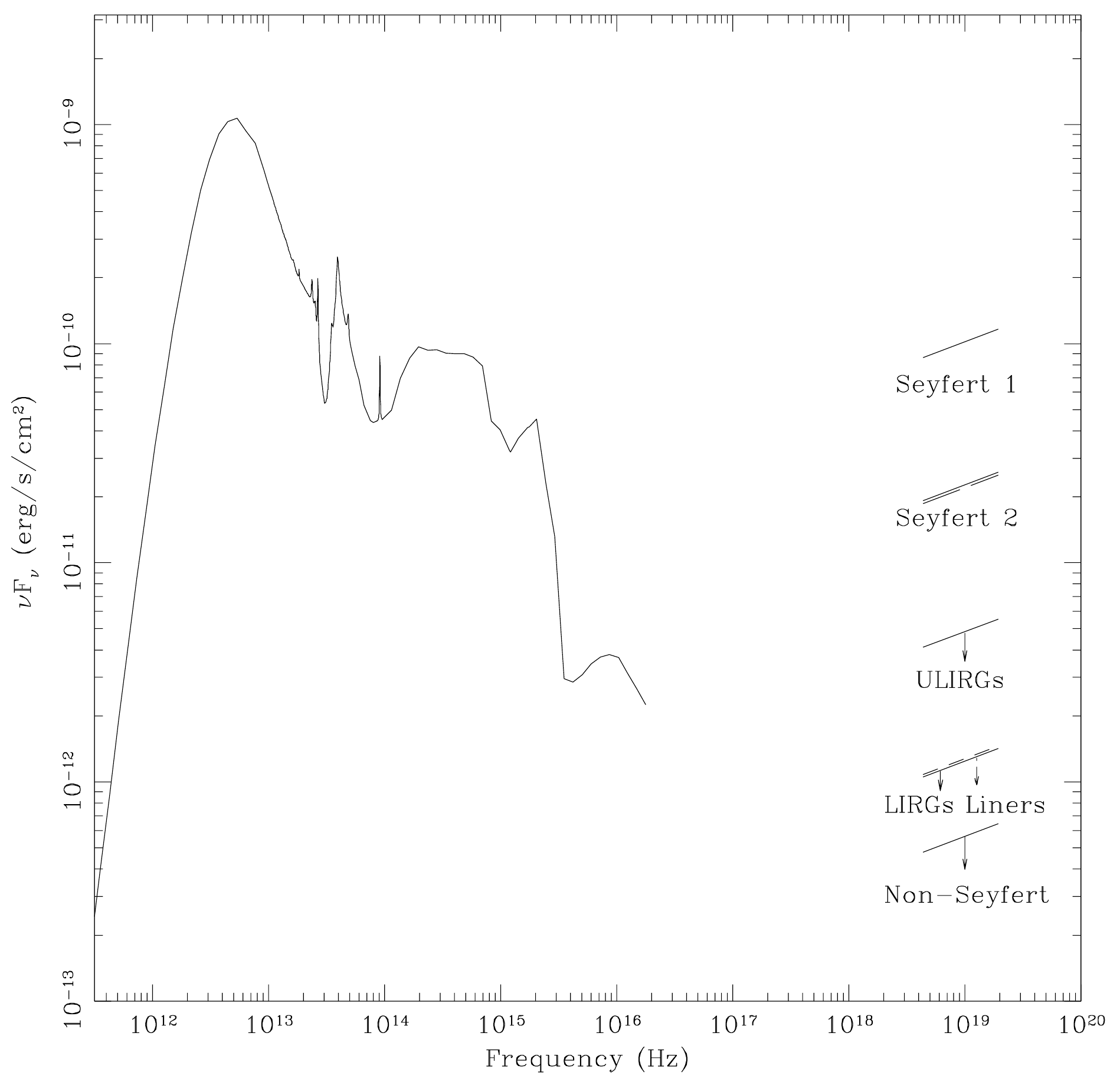}}
\caption{Average spectral energy distribution (SED) for the various galaxy subsamples, renormalized to an arbitrarily fixed 60 $\mu$m flux. 
The average ULIRGs infrared-optical SED is represented as a thin line for illustration. The hard X-ray spectral shape assumes a spectral slope of $\Gamma=1.8$. The long dash curve shows the relative hard X-ray contribution of Seyfert 1 galaxies, excluding NGC 4151.}
\label{fig:sed}
\end{figure}

At 60 $\mu$m most of the the emission is believed to originate from molecular clouds or diffuse 
dust reprocessing stellar light or from nuclear activity for luminous AGNs.
Figure \ref{fig:sed} presents the re-normalized average spectral energy distributions of each subsamples, 
together with the average infrared-optical spectra obtained for a sample of 30 ULIRGs \citep{Vega2008} 
for illustration. The spectra of the galaxy subsamples were re-normalized at 60 $\mu$m to the average 
obtained for the 30 ULIRGs.

The ratios between the flux densities at hard X-rays and at 60 $\mu$m are 12\% for Seyfert 1, 
3\% for Seyfert 2 (and for Seyfert 1 excluding NGC 4151), below 0.6\% for ULIRGs, below 0.1\% for the LIRGs and Liners, and finally
below 0.07\% for the ``non Seyfert" galaxies all together.

From the hard X-ray point of view we find a clear separation between the average Seyfert dominated 
and non Seyfert galaxies, the latter being not detected. As the hard X-rays are not much affected by moderate
absorption this difference could be intrinsic to the source samples and/or signature of Compton-thick 
absorption in the galaxy subsamples lacking Seyfert signatures.

The ratio between the flux densities obtained at hard X-rays from high-mass X-ray binaries and in the 
far infrared can be estimated from \cite{Grimm2003} and \cite{Kennicutt1998} and is not larger than 
0.03\%. This is two orders of magnitude lower than the hard X-ray signal obtained for the Seyfert subsamples 
and still below the upper-limits obtained for the ``non Seyfert" galaxies. Our upper-limits are therefore consistant 
with the hard X-ray emission expected from stellar activity. 
The ratio of the 60 $\mu$m to 18-80 keV flux densities in NGC 1068 \citep[$\approx500;$][]{Matt1997} is also 
smaller than the one obtained on average for the ``non Seyfert'' galaxies, in agreement with a lack of reflected 
hard X-rays in the latter.

The Seyfert 2 are 4.5 times more numerous than the Seyfert 1 in the RGBS in agreement with the results
of \cite{LaFranca2005} and \cite{Akylas2006} for a sample dominated by low luminosity sources. 
The hard X-ray emission of moderately absorbed objects is, at maximum, a factor of two lower than 
that of unabsorbed objects \citep[e.g.][]{Gilli2007}.
When compared to their 60 $\mu m$ emission, local Seyfert 1 appears on average 5 times brighter 
at hard X-rays than Seyfert 2. As most of this difference can be explained by the 
presence of NGC 4151, the sample is too small to elaborate on a possible genuine difference between 
the average unabsorbed spectral energy distributions of Seyfert 1 \& 2.

The hard X-ray upper-limit obtained for the ULIRGs indicates that, on average, these sources 
emit less than 20\% of the average emission detected in Seyferts when normalized to their infrared emission. 
This is consistent with the fact
that less than 30\% of the ULIRGs harbor a significant $(>10\%)$ AGN contribution in the far infrared 
\citep{Armus2007,Vega2008} and with the predominance of their starburst emission even at soft X-rays 
\citep{Franceschini2003}. Note that as the number of nearby ULIRGs is small, our upper-limits are 
an order of magnitude above the extrapolation at hard X-rays of the XMM-Newton detections.

Active nuclei absorbed by Compton-thick material in ULIRGs or LIRGs could, in principle, remain 
completely hidden at hard X-rays. 
To accomodate our hard X-ray upper-limits and assuming that the active nuclei have the same average 
intrinsic spectral energy distribution than obtained for our sample of Seyfert galaxies, no more than 3 and 20\% 
of the hard X-ray emission should be transmitted on average in LIRGs and ULIRGs respectively. For the ULIRGs, 
this means that active nuclei, if they exist, should be obscured by ${\rm N}_{\rm H} > 10^{24}~{\rm cm}^{-2}$ and 
that no more than a few of them in the 12 sources sample should remain unabsorbed. 
For the LIRGs, the active nuclei, if present, should be obscured by ${\rm N}_{\rm H} > 5\times 10^{24}~{\rm cm}^{-2}$ 
and no more than a few out of 144 should be unobscured.  

The ``non Seyfert" galaxies are, on average, very faint at hard X-rays. The upper-limit we have obtained 
indicates that their hard X-ray emission is not driven by active nuclei brighter than ${\rm L}_{\rm X}\sim 10^{41} ~{\rm erg/s}$ 
on average. The presence of bright active nuclei absorbed by ${\rm N}_{\rm H} > 5\times 10^{24}~ {\rm cm}^{-2}$ 
with no more than 15 unobscured ones out of 520 sources can however not be ruled out.

The hard X-ray upper-limit obtained for the Liner galaxies is in agreement with active nuclei or starbursts of low X-ray 
luminosity $(\sim 10^{40}~{\rm erg/s})$ reported in these objects \citep{Gonzales2006}.

\subsection{Contributions to the hard X-ray background}

We can estimate the relative contributions of our subsamples of galaxies
to the locally emitted hard X-ray background (CXB). The cumulative hard X-ray flux of the Seyfert
galaxies is $1.8\times 10^{-9}$ and $3.1\times 10^{-9}$ erg s$^{-1}$ cm$^{-2}$ for the type 1 and 2 respectively 
\citep[slightly less than 1\% of the total CXB;][and references therein]{Gilli2007}. 
The cumulative hard X-ray flux of the non Seyfert galaxies is $<0.38\times10^{-9}$  erg s$^{-1}$ cm$^{-2}$.

Assuming that Seyfert 1 and 2 are optically thin and, respectively, mildly absorbed \citep[which is not completely accurate in general; e.g.][]{deRosa2008, Brightman2008}, the fraction of the locally emitted CXB, emitted by absorbed objects $({\rm N}_{\rm H} > 10^{22}~{\rm cm}^{-2})$ in our sample, is 64\% and the fraction emitted by potentially Compton-thick non Seyfert objects is $<7\%~(5\sigma)$. 

These fractions can be compared to the predictions of CXB synthesis models and in particular to these of \cite{Gilli2007}, which could be obtained through the on-line tool \texttt{POMPA}\footnote{http://www.bo.astro.it/$\sim$gilli/counts.html}. Our sample is flux limited in the infrared, the corresponding X-ray flux limit for an individual Seyfert galaxy is on the order of $10^{-11} ~{\rm erg~s^{-1}~cm^{-2}}$. For such a flux limited sample the CXB synthesis model indicates that absorbed and Compton-thick sources should contribute for 65\% and, respectively, 18\% of the locally emitted CXB at 30 keV. 

The fraction of the locally emitted CXB to be accounted for by absorbed sources agrees with the results of our observations. The fraction expected to be emitted by Compton-thick sources is however significantly larger than the upper-limits obtained for non-Seyfert galaxies, which are therefore unlikely to harbor Compton-thick active nuclei contributing significantly to the locally emitted CXB. 

\section{Conclusions}

We have derived average hard X-ray fluxes for subsamples of Seyfert and non Seyfert galaxies stacking \textit{INTEGRAL/ISGRI} images of 613 members of the IRAS Revised Bright Galaxy Sample.

The Seyfert 1 \& 2 subsamples are clearly detected at hard X-rays while none of the other galaxy subsamples is detected.
When compared to their 60 $\mu m$ emission, local Seyfert 1 are on average 5 times brighter at hard X-rays than 
Seyfert 2. This can be explained by the presence of NGC 4151 dominating the small subsample of Seyfert 1.

The upper-limit obtained on the ULIRGs indicates that on average these objects emit less than 20\% of the average Seyfert emission at hard X-rays. This is consistant with previous results suggesting that the majority of these objects do not harbor significant AGN activity.

On average, ``non Seyfert" galaxies do not contain any active nuclei brighter than $10^{41}~{\rm erg/s}$ at hard X-rays.
If they harbor Compton-thick AGNs, these should be highly absorbed $(> 5\times 10^{24}~ {\rm cm}^{-2})$  and no more than a few \% of them could remain unabsorbed. 

The RGBS accounts for less than 1\% of the CXB. 64\% of this is produced by absorbed $(>10^{22} ~{\rm cm}^{-2})$ sources, in very good agreement with CXB synthesis models. Non Seyfert galaxies contribute for less than 7\% $(5\sigma)$ and have not been detected statistically. These galaxies are unlikely to harbor the Compton-thick AGNs required to synthesize the locally emitted CXB. Local Compton-thick AGNs, if they exist as a population, should therefore be looked for among Seyfert 2 or infrared weak galaxies.

\begin{acknowledgements}
Based on observations with INTEGRAL, an ESA project with instruments and science data centre funded by ESA member states (especially the PI countries: Denmark, France, Germany, Italy, Switzerland, Spain), Czech Republic and Poland, and with the participation of Russia and the USA. 
This research has made use of the SIMBAD database, operated at CDS, Strasbourg, France.
This work was performed in the frame of a ``stage de Master 1" of the University of Bordeaux.
\end{acknowledgements}

\bibliographystyle{aa}
\bibliography{10882}
\end{document}